\documentclass[final,5p,times,twocolumn]{elsarticle}

\usepackage{graphicx}

\usepackage{amssymb}

 \usepackage{lineno}

\usepackage{psfrag}

\journal{Nuclear Instruments and Methods in Physics Research, A}

\begin{document}

\begin{frontmatter}



\title{REAS3: A revised implementation of the geosynchrotron model for radio emission from air showers}


\author[KITEKP]{M. Ludwig\corref{cor}}
\author[KITIK]{T. Huege}

\address[KITEKP]{Karlsruher Institut f\"ur Technologie, Institut f\"ur Experimentelle Kernphysik, Campus S\"ud, 76128 Karlsruhe, Germany}
\address[KITIK]{Karlsruher Institut f\"ur Technologie, Institut f\"ur Kernphysik, Campus Nord, 76021 Karlsruhe, Germany}

\cortext[cor]{Corresponding author: Marianne Ludwig $<$marianne.ludwig@kit.edu$>$}

\begin{abstract}

Over the past years, the freely available Monte Carlo-code REAS which simulates radio emission from air showers based on the geosynchrotron model,
was used regularly for comparisons with data. However, it emerged that in the previous version of the code, emission due to the variation of the
number of charged particles within an air shower was not taken into account. In the following article, we show the implementation of these emission
contributions in REAS3 by the inclusion of ``end-point contributions'' and discuss the changes on the predictions of REAS obtained by this revision.
The basis for describing radiation processes is an universal description which is gained by the use of the end-point formulation. Hence, not 
only pure geomagnetic radiation is simulated with REAS3 but also radiation due to the variation of the net charge excess in the air shower,
independent of the Earth's magnetic field. Furthermore, we present a comparison of lateral distributions of LOPES data with REAS3-simulated
distributions. The comparison shows a good argeement between both, data and REAS3 simulations.

\end{abstract}

\begin{keyword}
radio emission \sep extensive air showers \sep modelling and simulation \sep endpoint contribution \sep geosynchrotron model
\end{keyword}

\end{frontmatter}


\section{Introduction}\label{introduction}

In recent years, radio detection of cosmic ray air showers has been developed further. With radio detector arrays like LOPES \cite{Falcke05},
\cite{HuegeArena2010} and CODALEMA \cite{Ardouin05}, \cite{Ardouin09}, correlations
of the radio signal with air shower parameters are studied and the dominance of the geomagnetic emission contribution was verified. To study the
physics of cosmic rays using radio signals, detailed theoretical simulations are needed. Many approaches for the modelling of radio emission exist,
but presently, there are two major approaches, both of which are based on geomagnetic effects \cite{HuegeArena2008}. On the one hand the
geosynchrotron model as implemented in REAS developed by Huege et al. \cite{HuegeFalcke2003a},\cite{HuegeFalcke2005a},\cite{HuegeFalcke2005b},
\cite{HuegeUlrichEngel2007a}, and on the other hand the macroscopic geomagnetic radiation model (MGMR) of Scholten, Werner and
Rusydi \cite{ScholtenWernerRusydi},\cite{WernerScholten2008a}. So far, the two models made conflicting predictions for the radio emission of cosmic
ray air showers. Essentially, this could be seen in the different pulse shapes (unipolar for REAS2 and bipolar for MGMR) and the differences in the
frequency spectra for low frequencies (dropping to zero 
for MGMR and levelling off for REAS2). The details of these differences and a comparison of both models are discussed in
\cite{HuegeLudwigScholtendeVries}. It arised that in REAS2, radiation due to the variation of the number of charged particles in EAS was
not considered as it is the case for nearly all time-domain approaches as well. The reason for this missing contribution was a flaw in the 
implementation of the radiation process of the geosynchrotron radiation. In REAS3 \cite{LudwigHuege}, this flaw was solved. In the following
sections, the details of the implementation as well as the results are illustrated.


\section{General functionality of REAS}\label{structure}

\begin{figure}[htb]
\centering
\includegraphics[width=0.35\columnwidth]{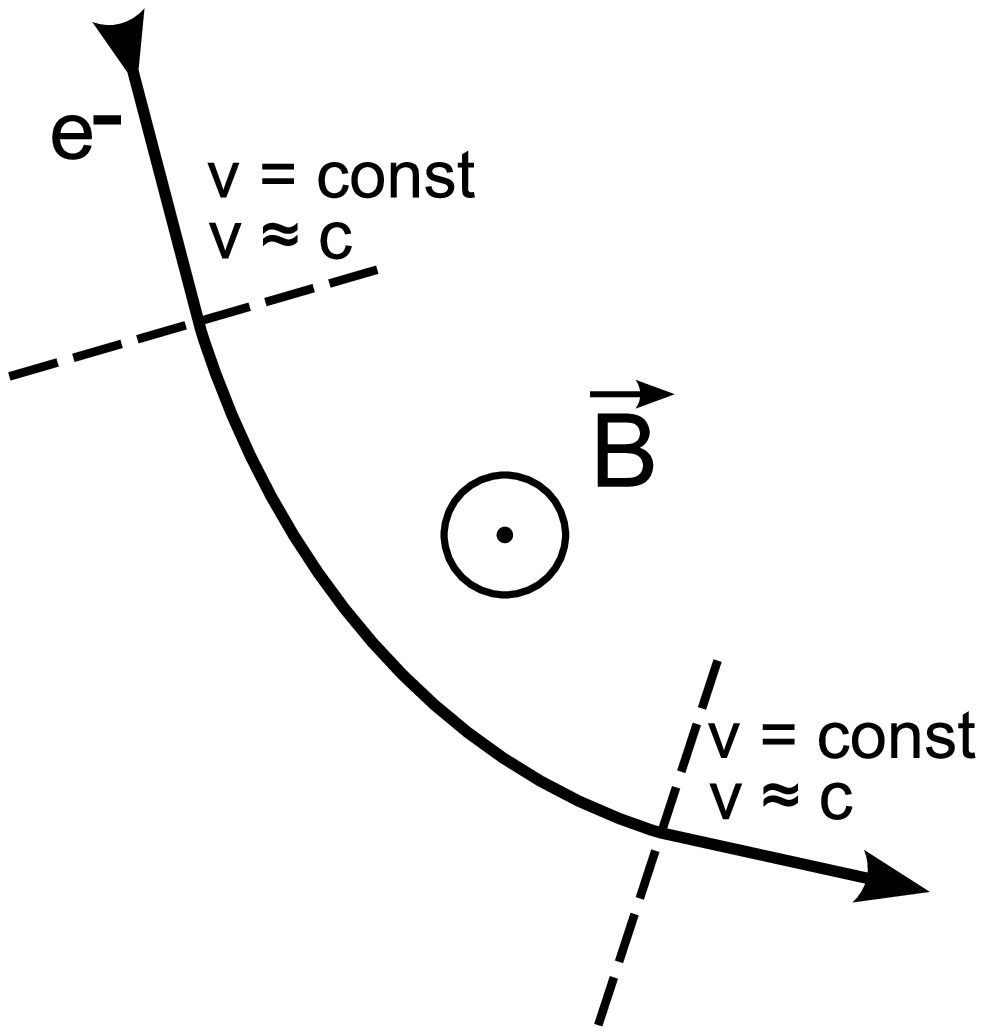} 
\hspace{0.5cm}
\includegraphics[width=0.4\columnwidth]{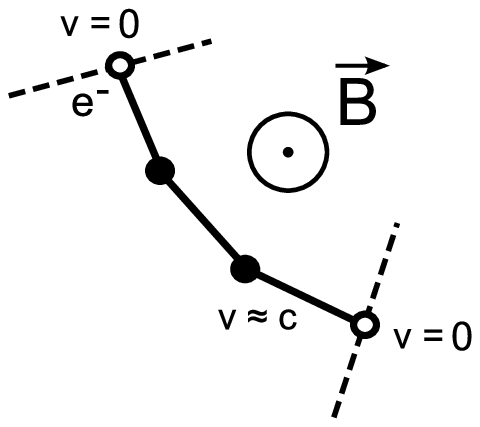}
\caption{\label{fig:tracks} Sketch of the trajectories how they are implemented in REAS. Left: REAS2. Right: REAS3.}
\end{figure}

To understand what was missing in the techincal implementation of radio emission in REAS2 it is helpful to know the general structure of this Monte
Carlo-code. First, the air shower is simulated with CORSIKA \cite{Corsika} saving all important information, e.g. the distribution of energy and
momentum of the particles, in histograms. On the basis of these histograms, in REAS, shower particles are generated according to the desired
distributions derived with CORSIKA. In the simulation code, each 
particle is followed analytically on its track through the Earth's magnetic field. Note that the real particle trajectories are described by several
unrelated short tracks. Finally, the radiation given from all shower particles is superposed for each single observer position. In REAS2, only
radiation processes along the trajectories were treated, but not at the end or the beginning of the tracks. This can be compared with a situation
that the particles arrive with velocity $v\approx c $ given by CORSIKA, enter the Earth's magnetic field where they are deflected on a short curved
track and finally fly out of the influence of the geomgagnetic field with velocity $ v\approx c $ (cf. left sketch of Fig. \ref{fig:tracks}). 
To revise the flaw in the derivation of the radiation process of the geosynchrotron radiation, emission contributions at the beginning and the end
of the tracks have to be taken into account, i.e. radiation due to the acceleration of the particle at the starting point of the trajectory and 
vice-versa deceleration of the particle at the stopping point is considered. To implement this
radiation, the best description of the radiation processes is an end-point formulation \cite{James}. In this case, the tracks are described by
straight track segments joined by ``kinks'' (cf. right sketch of Fig. \ref{fig:tracks}). If at a given atmospheric depth more particle trajectories
start than end, e.g. the number of particle declines, this results in a net contribution. 


\section{Incorporation of end-point contributions} \label{incorporation}

Adding the discrete end-point contributions to the continuous contributions along the tracks may produce problems, e.g., there is a risk of
double-counting. To get a consistent description of all radiation processes in the simulation, it is convenient to use the end-point formulation 
throughout. Radiation occurs if the velocity of the particle changes, i.e., in a kink of the track. Because the change of the velocity can be
considered instantaneous with respect to the times of interest ($\delta t \ll \frac{1}{\nu_{\mbox{\tiny{observed}}}}$ and
$\nu_{\mbox{\tiny{observed}}} \leq 100-1000\,$MHz) only the time-averaged process is of interest. Hence, the time-integrated field strength of the
radiation formula can be calculated. Equation \ref{eq:endpcalc} shows the result for the radiation in one kink of the track, 
\begin{eqnarray}
 \hspace{-0.5cm} \displaystyle\int\vec{E}(\vec{x},t)dt &=& \int_{t_1}^{t_2}\frac{e}{c} \left\vert \frac{ \vec{n}\times [(\vec{n}-\vec{\beta})\times \dot{\vec{\beta}}]} {(1-\vec{\beta}\cdot\vec{n})^3 R}\right\vert_{ret} \mathrm{d}t \nonumber \\
 & = & \vec{F}(t_2) - \vec{F}(t_1) \nonumber \\
 &=& \frac{e}{cR}\left(\frac{\vec{n}\times (\vec{n}\times \vec{\beta_2})} {(1-\vec{\beta_2}\vec{n})} \right) - \frac{e}{cR}\left(\frac{\vec{n}\times (\vec{n}\times \vec{\beta_1})} {(1-\vec{\beta_1}\vec{n})} \right) \label{eq:endpcalc}
 \end{eqnarray}
where $e$ indicates the particle charge,
$\vec{\beta} = \vec{v}(t)/c$ is given by the particle velocity, $R(t) = \vert\vec{R}(t)\vert$ describes the vector between particle and observer
position and $\vec{n}(t) = \vec{R}(t)/R(t)$ is the line-of-sight direction between particle and observer. The index ``ret'' means that the equation
needs to be	evaluated in retarded time. $ \vec{\beta_1} $ corresponds to the velocity before and $ \vec{\beta_2} $ to the velocity after the kink. In
this completly universal (cf.\,\cite{James}) and discrete calculation, radiation at the end or the beginning of the track corresponds to kinks where
one velocity is equivalent to zero, i.e., one term of the integrated sum vanishes. 


\section{Results} \label{comparison}

\subsection{Comparison between REAS2 and REAS3}

\begin{figure*}[hbt]
\centering
\includegraphics[angle=270,width=0.49\textwidth]{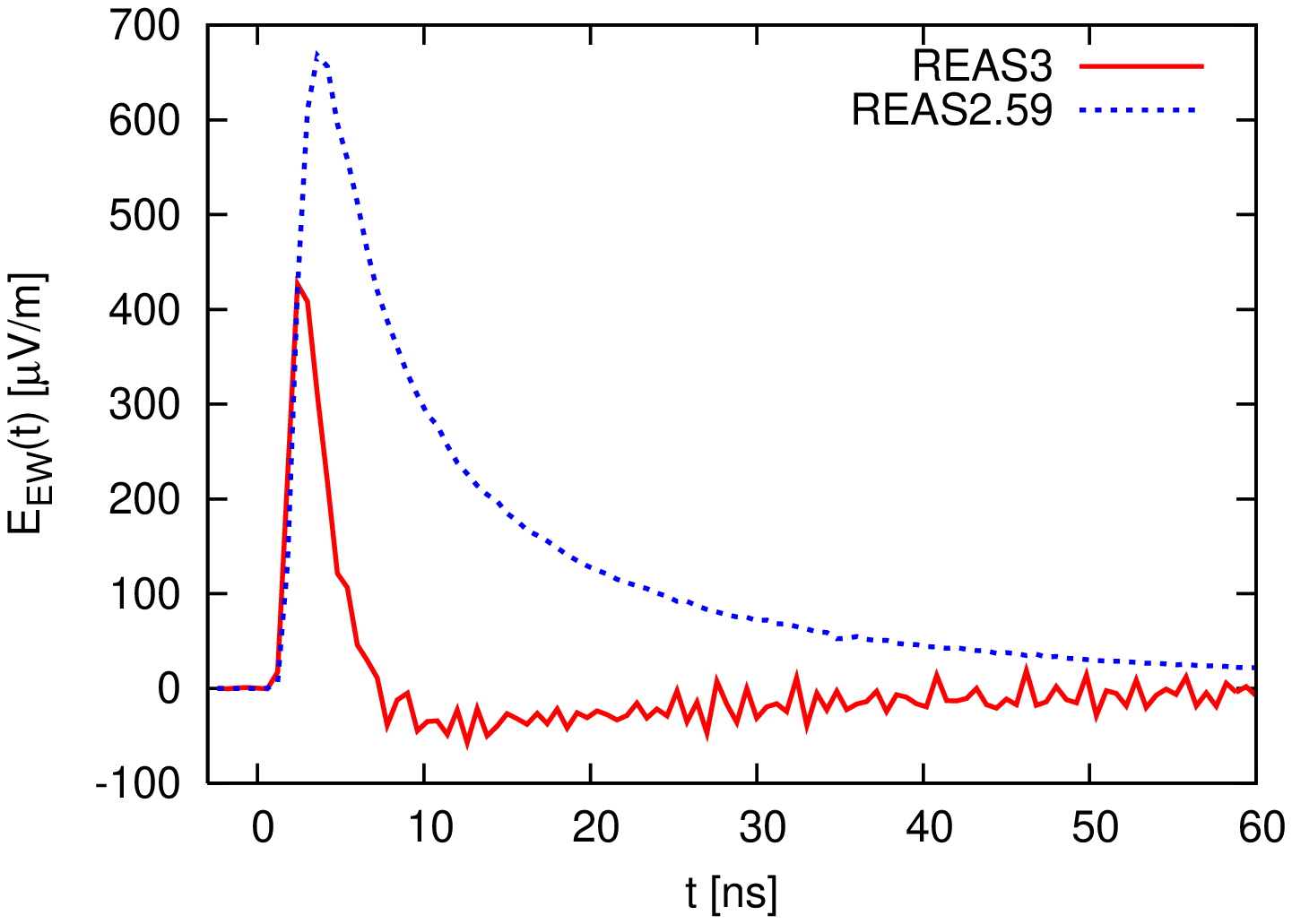}
\includegraphics[angle=270,width=0.49\textwidth]{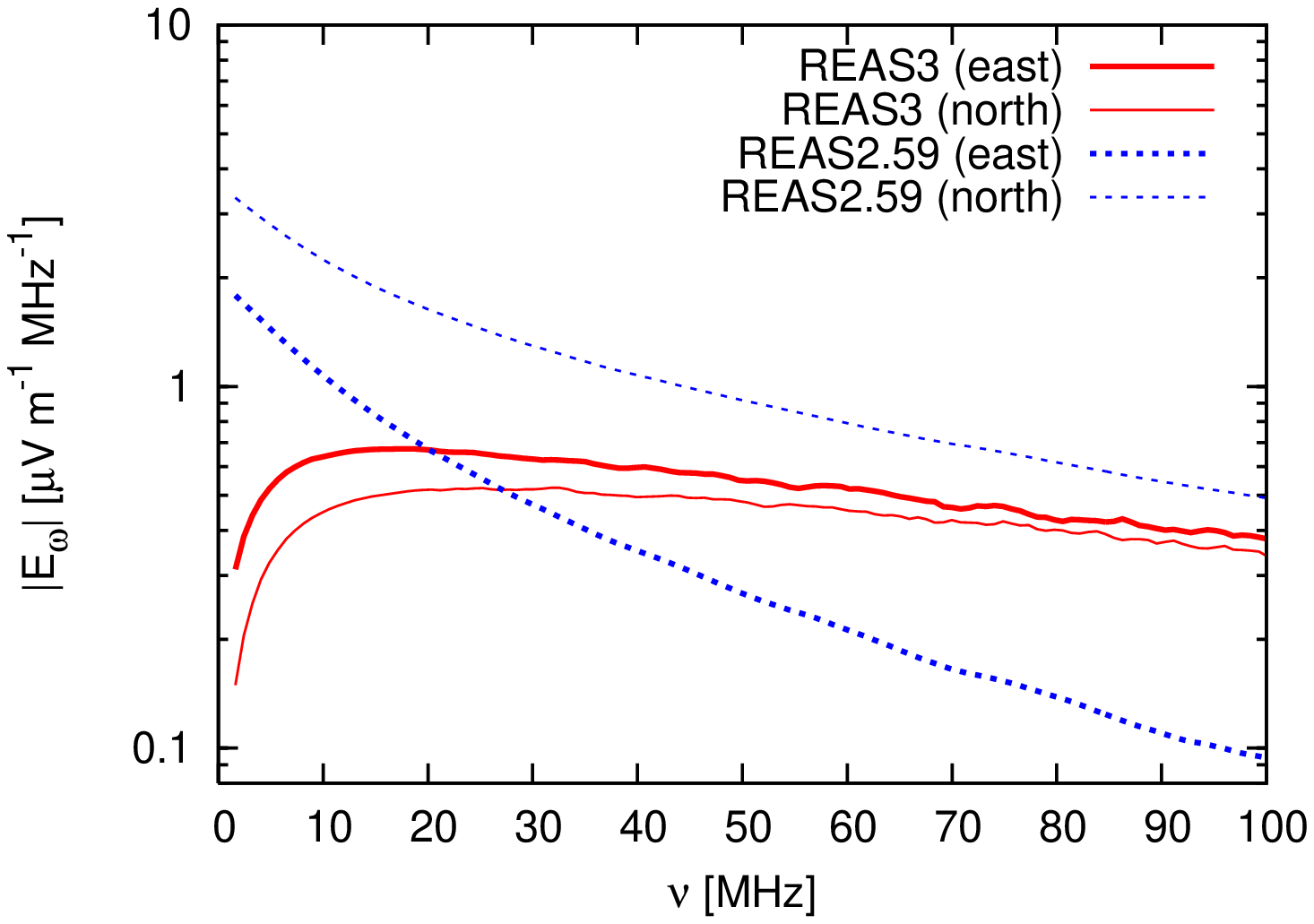}
\caption{\label{fig:R2vsR3} Direct comparison of REAS2 (dashed blue) and REAS3 (solid red) for a vertical air shower and observer distance of 100\,m. Left: Raw
pulse for an observer 100m north of the shower core. Right: Frequency spectra for observers east (thick lines) and north (thin lines) of the shower core.}
\end{figure*}

\begin{figure}[htb]
\centering
\includegraphics[angle=270,width=0.56\columnwidth]{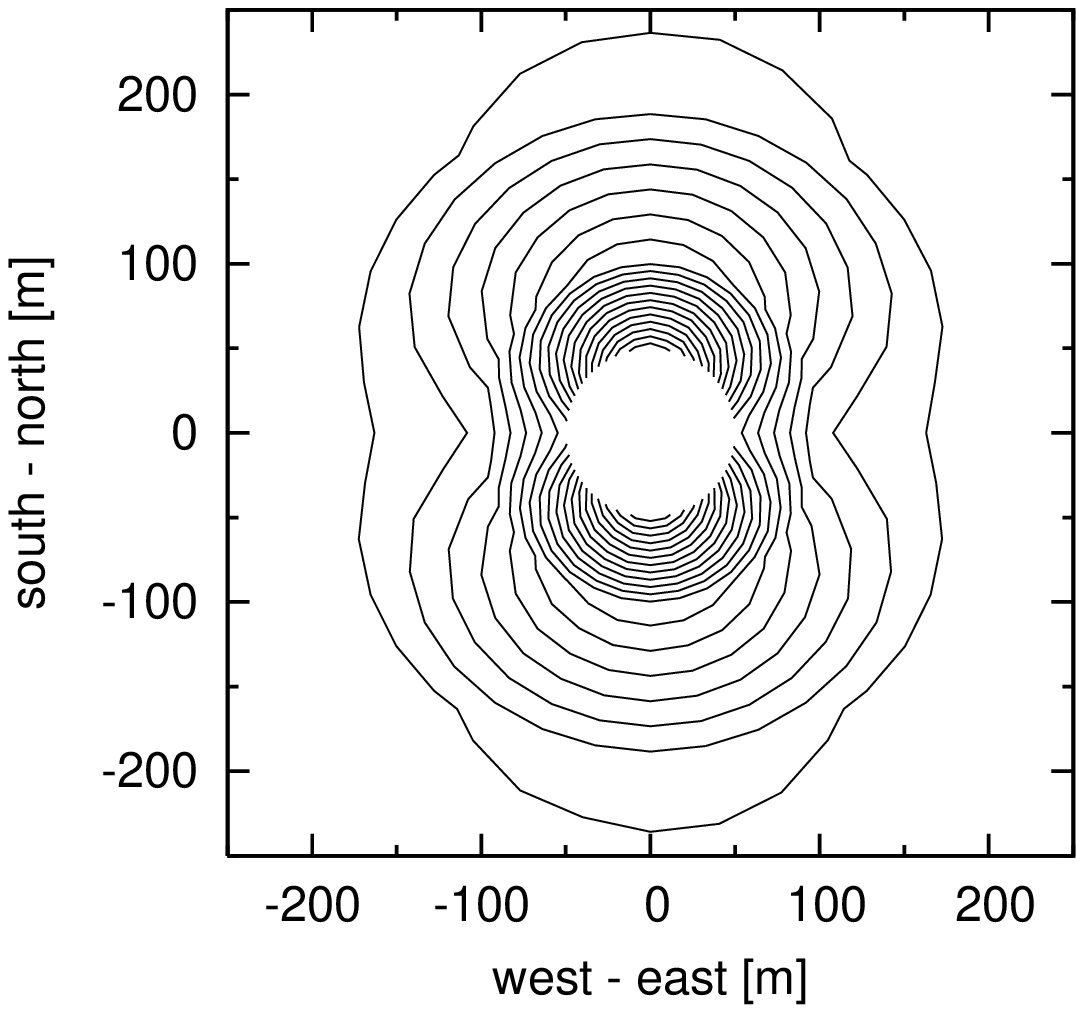}
\hspace{-1.2cm}
\includegraphics[angle=270,width=0.56\columnwidth]{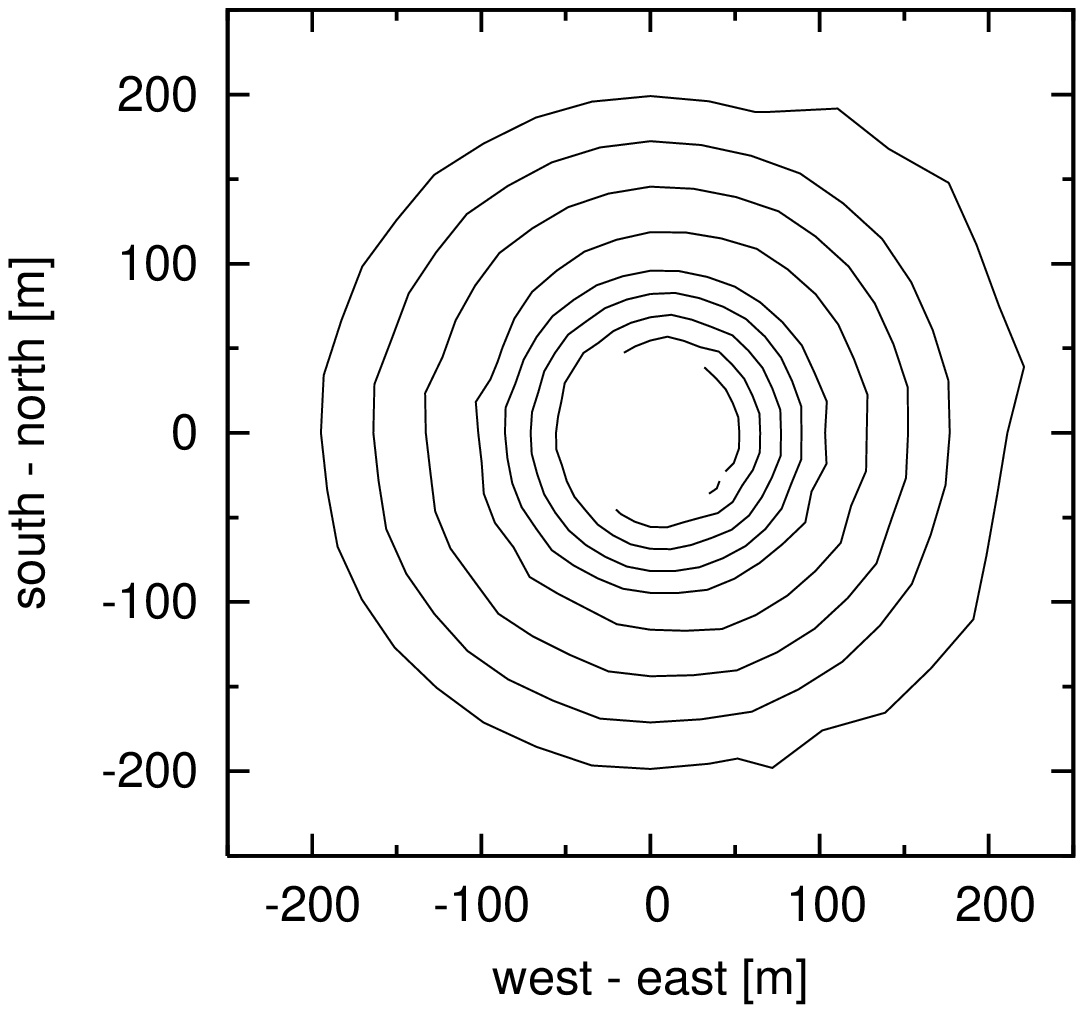}
\hspace{-1.2cm}
\caption{\label{fig:contour} Contour plots of the 60\,MHz absolute field strength for the radio emission from a vertical air shower. Left: 
REAS2. Right: REAS3. The white region in the center has not been simulated. Contour levels are 0.1\, $\mu$Vm$^{-1}$MHz$^{-1}$}
\end{figure}

In this section, a short overview about the major changes from REAS2 to REAS3 is given. In \cite{LudwigHuege} more details and discussion can be 
found. For the comparison, several simulations were done with a set of prototype showers. For this article, a simple shower geometry is chosen where
 the
geomganetic angle is 90$ ^\circ $, i.e., a vertical shower with a primary energy of $10^{17}\,$eV and a horizontal magnetic field of 0.23\,Gauss was
selected. Since one typical shower out of many CORSIKA simulated air showers was chosen, shower-to-shower fluctuations do not influence this
comparison. For REAS2 and REAS3 the same CORSIKA shower was taken as the basis. In Fig.\ref{fig:R2vsR3} the raw pulses of REAS2 and REAS3 for an
observer 100\,m north of the shower core are shown as well as the frequency spectra for observers 100\,m north and east of the shower core. 
It is obvious that the pulse shape changed from unipolar to bipolar. This change agrees with the theoretical expectation since the source of the
radio emission exists only over a finite time in a finite region of space (cf. \cite{ScholtenWernerArena2008}). In the frequency spectra (right 
plot of Fig. \ref{fig:R2vsR3}) this behaviour can be seen
as well because the spectral field strengths drop to zero for frequency zero. In addition, the spectral field strength for observers at different 
azimuthal positions differs less in REAS3 than in REAS2, which indicates an increased azimuthal symmetry of REAS3 compared to REAS2. For the spectra
of Fig. \ref{fig:R2vsR3} one observer in the north and in the east of the shower core is selected. In general, the spectra got flatter for REAS3.
The increased azimuthal symmetry is again visible in the contour plots of Fig. \ref{fig:contour} for the total field strength which gives an overall
impression of the changes from REAS2 to REAS3. 
In the contour plot of REAS3, an east-west asymmetry is distinguishable, i.e., the signal in the east is higher than the signal in the west. This is
explainable with the fact that in an air shower a charge excess of electrons occurs. The variation of the net charge excess leads to a 
further emission contribution which is discussed in the following section. 

\subsection{Discussion of charge excess}

\begin{figure}[htb]
\centering
\includegraphics[angle=270,width=0.56\columnwidth]{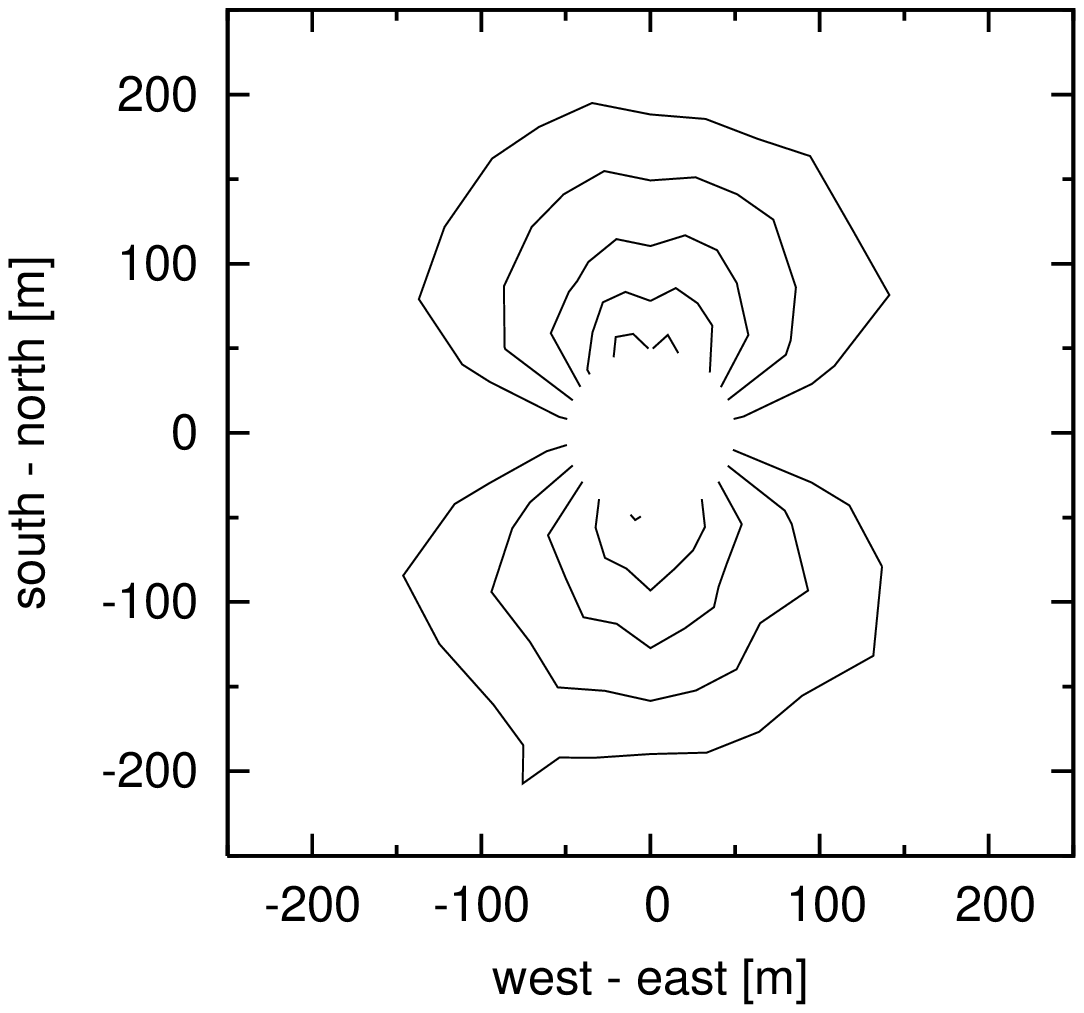}
\hspace{-1.2cm}
\includegraphics[angle=270,width=0.56\columnwidth]{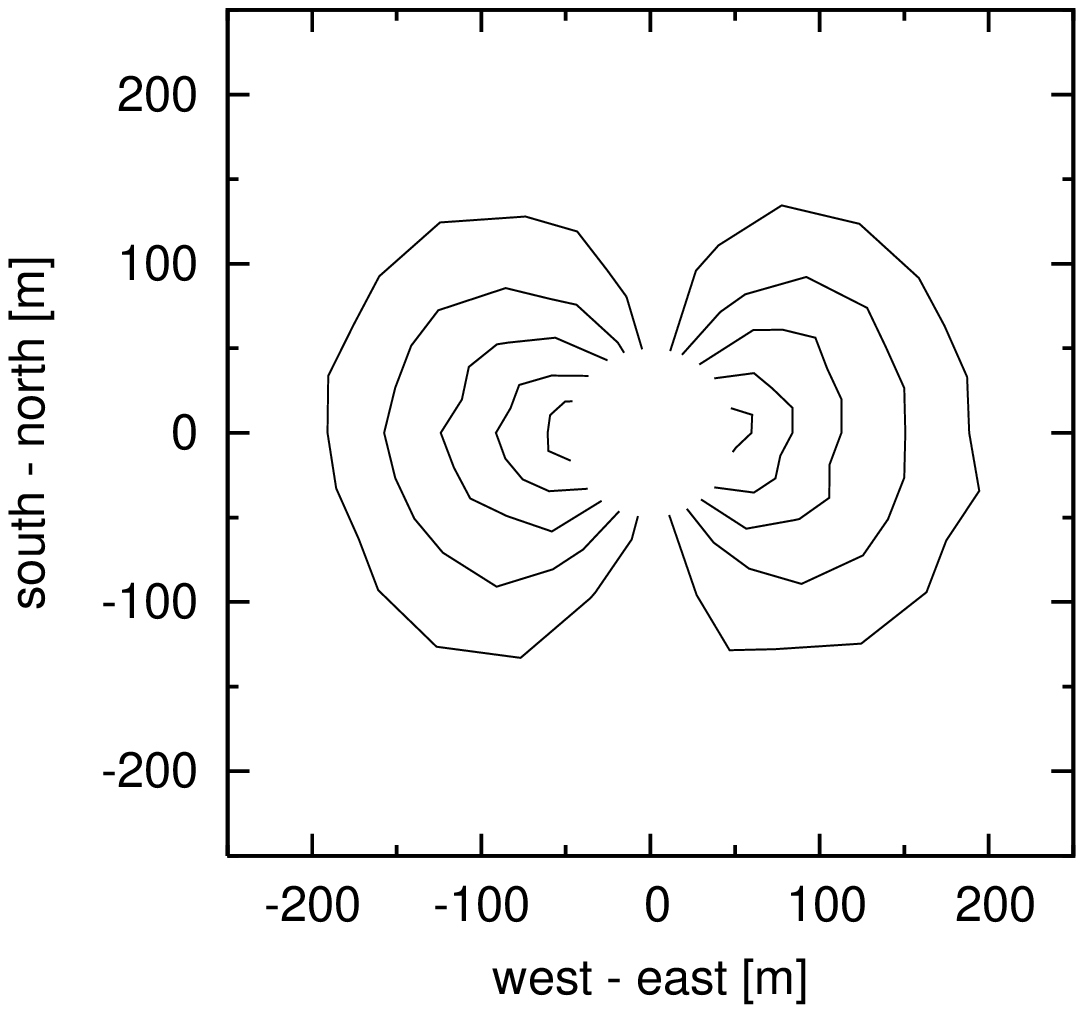}
\hspace{-1.2cm}
\caption{\label{fig:contour_ce} Contour plots of the 60\,MHz field strength for the charge excess component from a vertical air shower. Left: north polarization component.
Right: west polarization component. The white region in the center has not been simulated. Contour levels are 0.03\, $\mu$Vm$^{-1}$MHz$^{-1}$}
\end{figure}

Due to the time variation of the net charge excess in air showers, radio emission occurs even in the absence of any magnetic field. The same effect
appears
if the incoming direction of the air shower is parallel to the Earth's magnetic field. In REAS2, there was no radio emission for both cases because
radio emission arised only due to the deflection in the magnetic field. For a pure emission due to variation of the net charge excess a radially
polarised component is expected. To verify this, the radio emission of a vertical air shower with primary energy of $10^{17}\,$eV was simulated while
the strength of the magnetic field was set to 0\,Gauss. The contour plots of the 60\,MHz field strength for the north and the west polarization 
components are shown in Fig. \ref{fig:contour_ce}. The emission pattern is, as expected, radially polarised but the relative field strength of the 
charge excess emission at 60\,MHz is small in the distance range up to 200\,m compared to the relative field strength of total radio emission given
in Fig.\ref{fig:contour} (the contour levels for the simulation in this section are a factor of $ \sim $3.3 smaller than for the simulation with 
a realistic magnetic field).
Due to the charge excess in air showers and its variation it is evident that in the footprint of the radio signal a small azimuthal asymmetry has to
remain. Since the charge excess leads to a non-geomagnetic contribution in the radio emission, it is obvious that the radio emission of extensive air
showers is not purley $\vec{v}\times\vec{B}$ dependent.

\subsection{Comparison with LOPES data}

\begin{figure*}[htb]
\centering
\includegraphics[angle=270,width=0.49\textwidth]{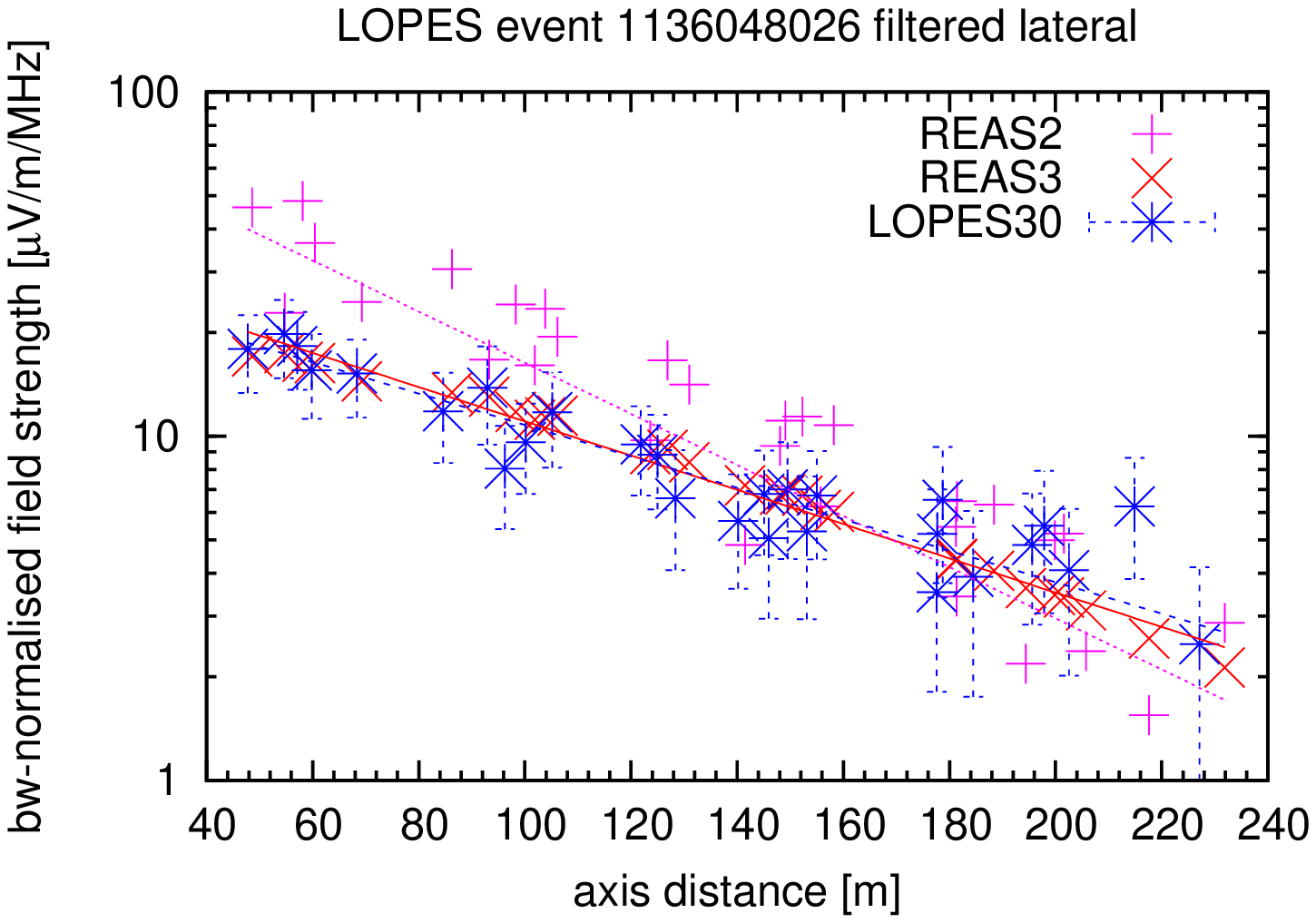}
\includegraphics[angle=270,width=0.49\textwidth]{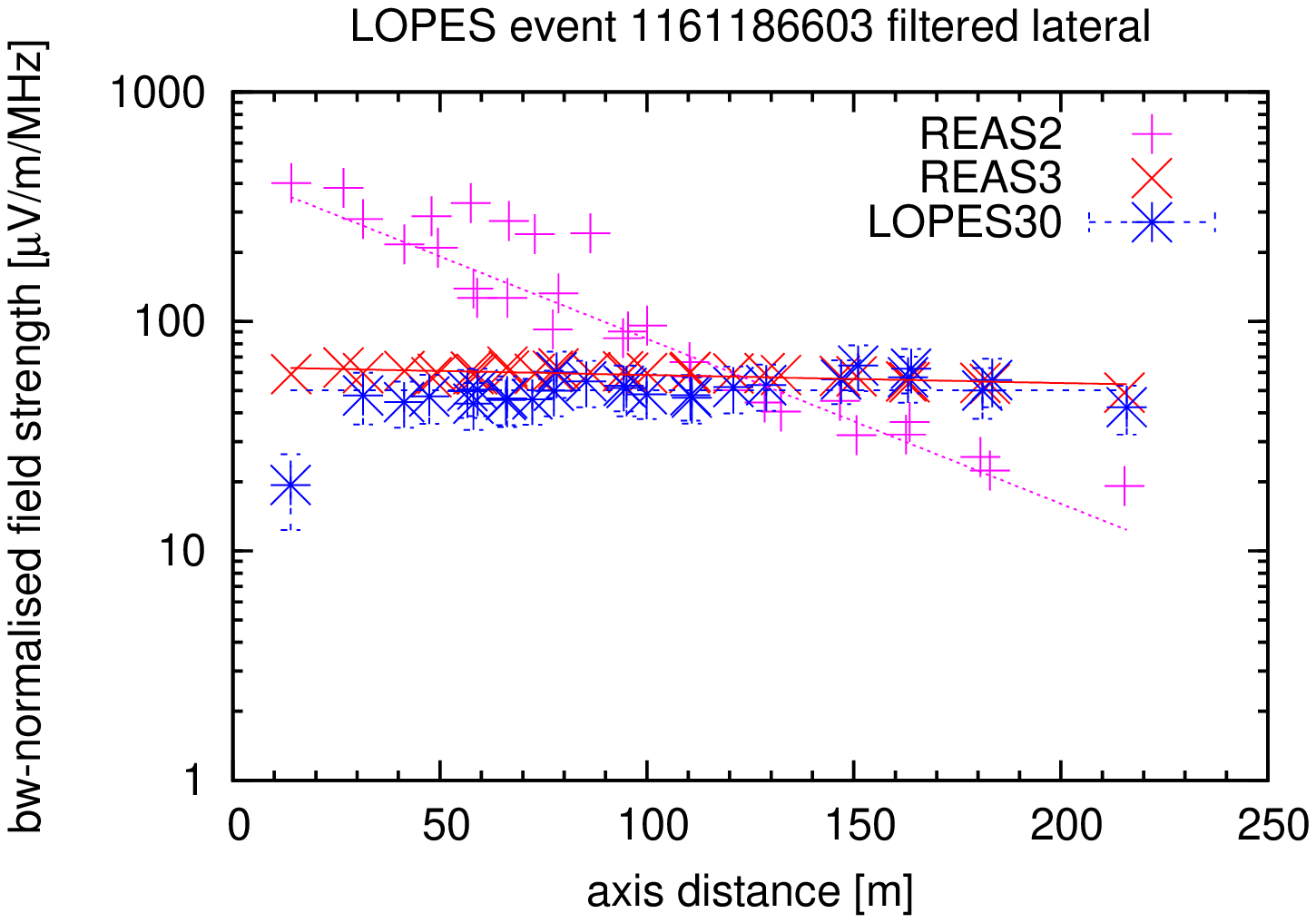}
\caption{\label{fig:lopes} Lateral distributions of two different LOPES events (blue $\ast$) compared with REAS2 (pink $ + $) and REAS3 (red 
$ \times $) simulations of proton induced air showers.}
\end{figure*}

Finally, to test the predictive capability of the REAS3 simulation on real data, it is suitable to compare the output of REAS3 and REAS2
with measured data. In this case, the lateral distributions of LOPES data have been used (published in \cite{Apel}). 
First, a set of 200 showers per each event was simulated with CONEX. Then, a ``typical'' shower was chosen out of this set to avoid shower-to-shower
fluctuations. For the selection, the averaged depth of the shower maximum $X_{\mathrm{max}}$ of all 200 showers was calculated and then a shower was
chosen with an $X_{\mathrm{max}}$ similar to the averaged one. For REAS2 and REAS3 the same CORSIKA shower was used to simulate the radio emission. 
In Fig.\ref{fig:lopes} two example events are shown for showers with proton as a primary particle. 
The LOPES event shown in the left plot has a  primary energy of $2.9\cdot 10^{17}$\,eV and a zenith angle of 31$^{\circ}$. The REAS3 simulation
reproduces the measured data very well. The second event has a primary energy of $2.9\cdot 10^{18}$\,eV while the zenith angle is 58$^{\circ}$. For
the first time, it is possible to reproduce also some flat lateral distributions which account for $\sim10\%$\, of all measured events. In general,
the lateral distribution with REAS3 became flatter than with REAS2 and agrees much better with the data. There are still some events which are not 
reproduced correctly by REAS3 but for all events the simulations are much closer to the measured data than REAS2 simulations. For future
investigations, the selection of a typical shower will be changed such that the ratio of the electron and muon number is comparable to the number
given by KASCADE instead of the selection via the $X_{\mathrm{max}}$. This could increase the agreement between simulations and measurements as well
as the noise correction \cite{Schroeder} on the data. For further comparisons between simulations and data a detector simulation, which is not
included so far, should be implemented to take into account the complete influence of the detector characteristics on the measured signal. 
Since REAS3 (the same is true for REAS2) has no free parameters, the
simulations are only dependent on the input parameters given by the shower characteristics such as primary mass, energy and incoming direction. 

\section{Conclusions}

The incorporation of end-point contributions as described in the article, led to a revision of the implementation of the geosynchrotron model
in the Monte Carlo-code REAS3. Due to the universal end-point formalism not only emission contributions due to the variation of the number of
charged particles in an air shower are taken into account but also contributions due to the deflection of the particles in the geomagnetic 
field as well as a radiation component arising due to the variation of the net charge excess in an extensive air shower. With this revision,
the pulse shape became bipolar and the frequency spectra are dropping to zero for zero frequency. Remaining asymmetries in the nearly symmetric 
azimuthal emission pattern are explainable by the time-varying charge excess in air showers. This results in a radio emission which is not purely of 
geomagnetic origin. REAS3 shows a good agreement between simulation and LOPES data comparing the lateral slopes, which is particularly striking 
since REAS3 has no free parameters at all. REAS3 is the first self-consistent time-domain model which takes the full complexity of air shower physics
as provided by CORSIKA into account. The code will be freely available for users worldwide. 

\section*{Acknowledgements}

The authors would like to thank many colleagues for very helpful discussions, in particular S.\ Buitink, R.\ Engel, H.\ Falcke, A.\ Haungs, C.W.\ James, 
O.\ Scholten and K.D.\ de Vries. This research has been supported by grant number VH-NG-413 of the Helmholtz Association.





\bibliographystyle{model1-num-names}

\begin{thebibliography}{25}
\expandafter\ifx\csname natexlab\endcsname\relax\def\natexlab#1{#1}\fi
\providecommand{\bibinfo}[2]{#2}
\ifx\xfnm\relax \def\xfnm[#1]{\unskip,\space#1}\fi

\bibitem{Falcke05} 
H.~{Falcke} et al. - LOPES Collaboration, Nature 435 (2005) 313--316.

\bibitem{HuegeArena2010} 
T.~{Huege} et al. - LOPES Collaboration, these proceedings (2010) arXiv:1009.0345.

\bibitem{Ardouin05} 
D.~{Ardouin} et al. - CODALEMA Collaboration, Nuclear Instruments and Methods in Physics Research A 555 (2005) 148--163.

\bibitem{Ardouin09} 
D.~{Ardouin} et al. - CODALEMA Collaboration, Astroparticle Physics 31 (2009) 192-200.

\bibitem{HuegeArena2008}
T.~{Huege} - Proc. of the ARENA 2008 conference, Rome, Italy, Nuclear Instruments and Methods in Physics Research A 604 (2009) 57--63.

\bibitem{HuegeFalcke2003a}
T.~{Huege}, H.~{Falcke}, Astronomy \& Astrophysics 412 (2003) 19--34.

\bibitem{HuegeFalcke2005a}
T.~{Huege}, H.~{Falcke}, Astronomy \& Astrophysics 430 (2005) 779--798.

\bibitem{HuegeFalcke2005b}
T.~{Huege}, H.~{Falcke}, Astroparticle Physics 24 (2005) 116.    

\bibitem{HuegeUlrichEngel2007a}
T.~{Huege}, R.~{Ulrich}, R.~{Engel}, Astroparticle Physics 27  (2007) 392--405.

\bibitem{ScholtenWernerRusydi}
O.~{Scholten}, K.~{Werner}, F.~{Rusydi}, Astroparticle Physics 29 (2008) 94--103.

\bibitem{WernerScholten2008a}
K.~{Werner}, O.~{Scholten}, Astroparticle Physics 29 (2008) 393--411.

\bibitem{HuegeLudwigScholtendeVries}
T.~{Huege}, M.~{Ludwig}, O.~{Scholten}, K.D.~{deVries}, these proceedings (2010) arXiv:1009.0346.

\bibitem{LudwigHuege}
M.~{Ludwig}, T.~{Huege}, submitted to Astroparticle Physics (2010).

\bibitem{Corsika}
D.~{Heck} et al., FZKA Report 6019, Forschungszentrum Karlsruhe (1998).

\bibitem{James}
C.W.~{James}, H.~{Falcke}, T.~{Huege}, M.~{Ludwig}, submitted to Phys Rev.\ E (2010) arXiv:1007.4146.

\bibitem{ScholtenWernerArena2008}
O.~{Scholten}, K.~{Werner} - Proc. of the ARENA 2008 conference, Rome, Italy, Nuclear Instruments and Methods in Physics Research A 604 (2009) 24--26.

\bibitem{Apel}
W.D.~{Apel} et al. - LOPES collaboration, Astroparticle Physics 32 (2010) 294-303.

\bibitem{Schroeder}
F.G.~{Schr\"oder} et al. - LOPES collaboration, these proceedings (2010).





\end{thebibliography}



\end{document}